\documentclass[a4paper,11pt]{article}
\usepackage[top=2cm,bottom=2.2cm,left=2.5cm,right=2.5cm]{geometry}
\usepackage{graphicx}
\usepackage{longtable}
\usepackage{hyperref}
\usepackage{natbib}
\usepackage[nottoc,numbib]{tocbibind}
\usepackage{authblk}
\usepackage{etex}
\usepackage{amssymb}
\usepackage{array}
\usepackage{mathtools}
\usepackage{url}
\usepackage{verbatim}

\usepackage[inline,shortlabels]{enumitem} 

\usepackage{ctable}
\setupctable{%
  botcap, 
  captionskip=2pt, 
  mincapwidth=2in, 
}

\usepackage[delarray]{tabu}

\newcommand{\param}[1] {\medskip \noindent \textbf{#1}~}
\newcommand{\paras}[1] {\smallskip \noindent \textbf{#1}~}

\newcommand{\lb} {\linebreak}

\newcommand{\otoprule} {\midrule[\heavyrulewidth]}

\newcommand{\ul}[1] {\underline{#1}}

\newfont{\helvetica}{phvr8t at 10pt}
\newfont{\helveticasmall}{phvr8t at 9pt}
\newfont{\helveticabig}{phvr8t at 22pt}
\newfont{\helveticao}{phvro8t at 11pt}

\ifdefined\labelenumiii
  \renewcommand{\labelenumiii}{(\roman{enumiii})}
\else
  \newcommand{\labelenumiii}{(\roman{enumiii})}
\fi

\title{Exploring Query Categorisation for Query Expansion: A Study}
\author[1]{Dipasree Pal \thanks{dipasree.pal.gmail.com; Corresponding author. Fax.: +91 33 2577 3035; Tel.: +91 33 2575 2858. }}
\author[1]{Mandar Mitra \thanks{mandar.mitra@gmail.com; Fax.: +91 33 2577 3035; Tel.: +91 33 2575 2858.}}
\author[2]{Samar Bhattacharya \thanks{samar\_bhattacharya@ee.jdvu.ac.in; Fax.: +91 33 2577 3035; Tel.: +91 33 2414 6129}}

\affil[1]{Indian Statistical Institute, 203 B.T. Road, Kolkata-700108, India}
\affil[2]{Jadavpur University, 188, Raja SCM Road, Kolkata-700032, India}
\date{}
\begin{document}

\maketitle

\begin{abstract}
The vocabulary mismatch problem is one of the important challenges facing
traditional keyword-based Information Retrieval Systems. The aim of query
expansion (QE) is to reduce this query-document mismatch by adding related
or synonymous words or phrases to the query. 

Several existing query expansion algorithms have proved their merit, but
they are not uniformly beneficial for all kinds of queries. Our long-term
goal is to formulate methods for applying QE techniques tailored to
individual queries, rather than applying the same general QE method to all
queries. As an initial step, we have proposed a taxonomy of query classes
(from a QE perspective) in this report. We have discussed the properties of
each query class with examples. We have also discussed some QE strategies
that might be effective for each query category. 

In future work, we intend to test the proposed techniques using standard
datasets, and to explore automatic query categorisation methods.

\end{abstract}

\section{Introduction}
\label{cha:introduction}
The use of Search Engines (SEs) has become an inseparable part of the
activities of most computer users. People use SEs in various forms to find
information in a wide variety of contexts: from Web search through desktop
search and email search to searching through document archives belonging to
specific domains such as the medical and legal domains. Depending on the
information need, finding the desired information can be a more or less
difficult task.

The well-known \emph{vocabulary mismatch} problem is one significant factor
that makes searching difficult. A user's query $Q$ and a useful document
$D$ in a document collection may use different vocabulary to refer to the
same concept. Retrieval systems that rely on keyword-matching may not
detect a match between $Q$ and $D$. A good retrieval system must bridge the
potential vocabulary gap that exists between useful documents and the
user's query. Query Expansion (QE), the addition of related terms to a
user's query, is one important technique that attempts to solve this
problem by increasing the likelihood of a match between the query and
relevant documents.

Most lay users prefer to keep their interaction with a retrieval system
simple. Thus, most QE methods are completely automatic and involve little
or no additional effort on the part of a user. Of course, a completely
automatic QE method may end up adding unrelated terms to a user's query,
thus changing the query's focus. This is known as \emph{query drift}. In such
cases, QE causes performance to deteriorate rather than improve.

\begin{figure}[t]
  \centering
  \includegraphics[width=0.9\textwidth]{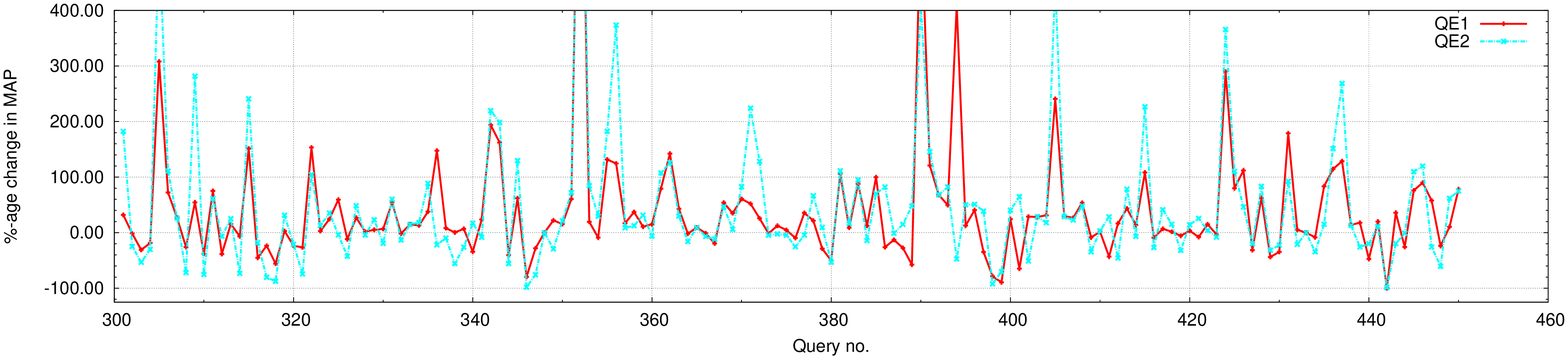}
  \caption{Variability of QE techniques across queries}
  \label{fig:variability}
\end{figure}

Over the years, many different QE techniques have been proposed. A recent
survey of such techniques can be found in~\citep{carpineto-survey}. While a
number of QE techniques have been shown to be effective on average
(i.e.\ when their overall impact across a large set of queries is measured),
the effect of different QE techniques for individual queries can vary
greatly. Figure~\ref{fig:variability} makes this point
graphically\footnote{Details about the dataset and the techniques used to
  generate these plots can be found in the Appendix.}. The points on the
X-axis represent individual queries; the Y-axis denotes the relative
improvement in performance obtained for each query by using query expansion.
The lines labelled QE1 and QE2 correspond to two different QE techniques. 
Points on QE1 (resp.\ QE2) that lie above the X-axis correspond to queries
for which this expansion method improves performance, while a point below
the X-axis corresponds to a query for which the method hurts retrieval
effectiveness.

Table~\ref{tab:variability} shows the Mean Average Precision (MAP) scores
for three retrieval methods: a baseline strategy that uses the original,
unexpanded queries, QE1 and QE2. QE1 and QE2 are clearly superior to the
baseline on average. This reinforces the claim above that QE techniques
often improve overall performance. However, it is clear from
Figure~\ref{fig:variability} that the impact of QE1 or QE2 varies greatly
across queries. Specifically, QE1 and QE2 result in
decreased performance for a number of queries. Also, while the overall
performance figures for QE1 and QE2 are comparable, each of these methods
outperforms the other on about half the queries used in this experiment.

The hypothetical performance that would be obtained if one could predict in
advance the most effective technique for a query --- no expansion vs.\ QE1
vs.\ QE2 --- is shown in the last column (MAX) of Table~\ref{tab:variability}. 
Notice that such a capability would lead to nearly 35\% improvement in
retrieval effectiveness.

\bigskip \medskip
\begin{table}[h]
 \centering
 \begin{tabular}{ccccc}\hline
        & Baseline & QE1                & QE2                & MAX    \\\hline
   MAP  & 0.1842   & 0.2191 (+ 18.95\%) & 0.2183 (+ 18.51\%) & \textbf{0.2473 (+ 34.26\%)} \\\hline
 \end{tabular}
 \caption{Potential improvement obtainable in principle by judiciously choosing QE techniques}
 \label{tab:variability}
\end{table}

In their overview of the NRRC Reliable Information Access (RIA) Workshop,
\cite{Harman2009} make a similar point: ``it may be
more important for research to discover what current techniques should be
applied to which topics, rather than to come up with new techniques''.

\subsection{Problem statement}
In this study, we consider the important problem of predicting the most
effective QE technique for a given query (including the possibility that
not expanding certain queries may be most effective). We explore one
possible approach to this question. We examine a number of different
criteria that can be used to classify queries. For each query category, we
discuss what QE techniques (or more generally, what query processing
techniques) might be most effective.

Our eventual goal is to find methods that can automatically (or
semi-automatically, i.e., with some assistance from a user) classify a
given query into one (or sometimes more) of several pre-defined categories.
We will then apply the QE method that is most appropriate for this
category. Our hypothesis, supported by Table~\ref{tab:variability} and
\cite{Harman2009}, is that overall performance should improve if we apply
QE techniques specifically tailored to a given query, rather than applying
the same general QE method to all queries.


\subsection{Outline}
\label{sec:outline}
The rest of this report is organised as follows.
Section~\ref{cha:related-work} discusses related work and its relationship
to this study. Section~\ref{cha:query-types} presents a taxonomy of query
categories. For each category, we provide examples of queries belonging to
that class. We also discuss QE techniques that are likely to be most
effective for that category. Details about the data presented in this
Section are given in Appendix~A. We conclude in
Section~\ref{cha:conclusions} by presenting a summary of the work done
along with a roadmap for future work.

\section{Related Work}
\label{cha:related-work}
Related work can be broadly classified into three categories: research
related to query categorisation, prior work on query expansion, and
research on query performance prediction.

\subsection{Query classification}
\label{sec:query-classification}
Automatic query categorisation (QC) is a well-known problem that has been
studied for many years in both the Information Retrieval and Machine
Learning communities. QC is usually treated as a multi-class categorisation
problem. It is quite different from normal text categorisation, since
queries are not as long as text documents.

Different types of query classification approaches have been defined
according to the purpose that the classification is intended to serve. A
well-known classification of Web
queries~\citep{DBLP:journals/sigir/Broder02} uses three categories:
informational, navigational, and transactional. Navigational queries are
entered by users looking for a specific website, whereas informational
queries cover a broad topic, for which there are typically many relevant
documents. Transactional queries have commercial / transactional purposes.
Transactional queries or queries with commercial intent are further
classified in \citep{commercial} depending on whether the user has
``on-line commercial intent'' (i.e.\ intention to purchase a product or
utilise a commercial service). Naturally, these categories are not
applicable to general-purpose queries that have no commercial intent. On a
somewhat related note, \cite{conf/spire/Baeza-YatesCG06} classify Web
queries according to whether they are informational, non-informational or
ambiguous.

Another traditional approach classifies queries according to the domain or
subject area targeted by the query. For example, the KDDCUP competition
2005\footnote{\url{http://www.acm.org/sigs/sigkdd/kdd2005/kddcup.html}}~\citep{kddcup}
focused on a Web query classification task. This task defined 67 query
categories organised into a hierarchical taxonomy, for example
\emph{Computers / Security}, \emph{Computers / Software},
\emph{Entertainment / Celebrities}, \emph{Sports / Tennis}. A single query
could belong to multiple categories. For example, relevant documents for
the query ``Beijing 2008'', may belong to the following domains:
\emph{Sports / Olympic Games}, \emph{Information / Local \& Regional},
\emph{Living / Travel and Vacation} and \emph{Information / Law and
  Politics}. Thus, this query belongs to multiple categories. Competitors
were required to classify 800,000 real user queries into the 67 categories.
Out of these queries, only 800 queries (randomly chosen) were labeled
manually, among which 682 queries belonged to multiple
categories~\citep{cao-kdd}.

\citet{Beitzel:2004:HAV:1008992.1009048} reported 16
categories of Web queries. These query classes are also based on the
subject domain of relevant documents, like music, games, entertainment,
computer, health, US-sites. The authors analysed Web traffic on an hourly
basis using these query types. They showed that music related queries cover
50\% of the total queries, while queries targeted at US-sites cover 35\%
queries, and queries related to entertainment comprise 5\% of the overall
query set.
\citet{DBLP:conf/cikm/CaoSXHYC09} also classify Web queries
into 17 groups with the aim of improving personalised search, but details
about these query groups are not available.

In recent times, query classification has become a particularly important
problem since most Web search engines earn their revenue via targeted
advertisements provided alongside search results.
\citet{Gabrilovich:2009:CSQ:1513876.1513877} classify queries onto a
fine-grained commercial taxonomy with approximately 6000 nodes, arranged in
a hierarchy with median depth 5 and maximum depth 9. The key idea in this
approach is to determine the class of a query by classifying the search
results retrieved for that query.

\bigskip 
Apart from the targeted domain, queries may also be classified based on
certain features, for example,
\begin{enumerate*}[(i)]
\item \emph{ambiguous} queries,
\item \emph{short} queries, and
\item \emph{hard} or \emph{difficult} queries.
\end{enumerate*}
\citet{xu-sigir09} classify queries into three categories, based on their
relationship with Wikipedia topics. These categories are: 
\begin{enumerate*}[(i)]
\item queries about specific entities;
\item ambiguous queries; and 
\item all other queries.
\end{enumerate*}
Hard queries were studied in the Robust Track at
TREC\footnote{\url{http://trec.nist.gov}}~\citep{robust}.
Substantial work has also been done on queries that have multiple aspects
~\citep{Harman:1988:TIQ:62437.62469,Buckley:2009:WCI:1644394.1644417}.
In the next section (Section~\ref{cha:query-types}), we discuss various criteria for query classification,
including some of the criteria mentioned above. While some of these query
types have been defined by other researchers in earlier work, we
specifically investigate the relationship between query categories and
appropriate QE strategies.


\subsection{Query performance prediction}
\label{sec:qpp}
Query performance prediction may be regarded as a special case of the query
categorisation problem, in which the objective is to classify a query as
being either hard or easy for a given retrieval system. \cite{pqp} were
among the earliest to study the Query Performance Prediction (QPP)
problem~. They defined the \emph{Clarity Score} as the relative entropy
between a query language model and the corresponding collection language
model. This score is intended to measure the ambiguity of a query with
respect to a document collection. The authors showed that the Clarity Score
is positively correlated with average precision (a standard evaluation
measure) on a variety of benchmark datasets.

QPP methods may be classified into two broad categories.
\begin{enumerate}
\item \emph{Pre-retrieval} methods. These methods (e.g., the method
  proposed in~\cite{he_query_2006}) use only the initial query, and term
  statistics from the target document corpus collected at indexing time. In
  particular, no preliminary retrieval results are needed.
  ~\citet{hauff_survey_2008} present a survey of pre-retrieval
  QPP methods. 
\item \emph{Post-retrieval} methods. These methods additionally make use of
  the results retrieved in response to the initial query, usually by
  analysing the similarity scores for the retrieved documents. The
  prediction method proposed by ~\citet{shtok:lncs,shtok:2012}
  may be regarded as a representative post-retrieval method. It is based on
  the hypothesis that ``high standard deviation of retrieval scores in the
  result list correlates with reduced query-drift, and consequently, with
  improved effectiveness.''
\end{enumerate}
A good introduction to work in this area can be found in a monograph
by~\cite{carmel_estimating_2010}. The monograph provides the background and
motivation for the QPP problem. It covers pre-retrieval and post-retrieval
methods, as well as methods that combine these two approaches. Finally, it
also discusses applications of query difficulty estimation. A more
up-to-date overview is provided in a tutorial by~\cite{carmel_query_2012}.

Recently, \cite{Kurland:2012:BRP:2396761.2396866} have proposed a
probabilistic framework for QPP that unifies various earlier, apparently
diverse approaches~\citep{pqp,VinayCMW06,Yom-Tov:2005,Zhou:2006,Zhou:2007}.
\cite{represent-qpp} generalise this framework by modelling
the user's actual information need (as represented by the query). Their
framework makes it possible to integrate pre-retrieval, post-retrieval, and
query-representativeness based predictors.

\subsection{Query expansion}
\label{sec:query-expansion}
A great deal of work has been done on QE. \citet{carpineto-survey} provide a comprehensive and up-to-date
survey of various automatic QE techniques. In earlier work on QE, we find
that \emph{expansion terms} (i.e.\ the terms that are added to the original
query) are generally selected from 3 types of sources. On the basis of the
source of expansion terms, QE strategies can be divided into the following
groups.

\begin{itemize}
\item \textbf{Local:} ``Local'' QE techniques select candidate expansion
  terms from a set of documents retrieved in response to the original
  (unexpanded) query. Ideally, expansion terms should be drawn from some
  initially retrieved \emph{relevant} documents. Since these documents are
  relevant, terms present in these documents are expected to be related to
  the query, and should help to retrieve other similar documents which are
  also likely to be relevant. If the user does not provide any feedback
  about which of the initially retrieved documents are relevant, certain
  simplifying assumptions may be made. Usually, in the absence of user
  feedback, a few top-ranked documents are assumed to be relevant. This is
  called \emph{pseudo relevance feedback} (PRF). This method has an obvious
  drawback: if several of the documents assumed to be relevant are in fact
  non-relevant, then the words added to the query (drawn mostly from these
  documents) are unlikely to be useful expansion terms, and the quality of
  the documents retrieved using the expanded query is likely to be
  poor.

  \citet{mitra-improving} propose a local expansion method that
  tries to prevent query drift by ensuring that the query is expanded in a
  balanced way. 
  \citet{lgca,lca} present a method called \emph{local context
    analysis} that also obtains candidate expansion terms from a few
  top-ranked documents. These terms are scored on the basis of their
  co-occurrence patterns with all of the query terms. The highest scoring
  terms are added to the query. 
  Recently, \cite{colace2015improving} have demonstrated the effectiveness
  of a new expansion method that extracts weighted word pairs from relevant
  or pseudo-relevant documents. 
  Researchers have also applied learning to rank methods to select useful
  terms from a set of candidate expansion terms within a PRF
  framework~\citep{xu2015assessment}.

\item \textbf{Global:} ``Global'' QE techniques select expansion terms from
  the entire database of documents. Candidate terms are usually identified
  by mining term-term relationships from the target corpus.
    
  \citet{Qiu-frei} propose a global QE technique that makes
  use of a \emph{similarity thesaurus}. A similarity thesaurus is a matrix
  containing term-term similarity scores as its entries. These similarity
  scores are computed based on how word-pairs co-occur in the documents
  contained in a corpus. Expansion terms are selected on the basis of a
  probabilistic measure of a term's relationship to the query concept.

  \citet{phrasefinder} also propose a global technique,
  called \emph{phrasefinder}, that is based on term co-occurrence
  information in the corpus. Each term $T$ corresponds to a vector $V_T$ of
  associated (or co-occurring) terms. A term $T$ is assigned a similarity
  score based on the similarity between the original query and $V_T$. The
  terms that are most similar to the query are selected as expansion terms.

  \citet{mult-db-gauch} define two words as similar if they
  occur in similar \emph{contexts}, where a word's context is defined in
  terms of its neighbouring words in a corpus. Words that are similar to
  the query words are selected for inclusion in the expanded query.

  \citet{carpineto-kld} use a combination of local and
  global approaches. Their hypothesis is that a useful term will occur
  more frequently in relevant documents than in non-relevant documents or
  in the whole corpus. \citet{long-span} also combine
  local and global information in the form of \emph{long-span collocates}
  --- words that significantly co-occur with query terms. Collocates of
  query terms are extracted from both the entire corpus, as well as from a
  subset of retrieved documents. The significance of association between
  collocates is estimated using modified Mutual Information and $Z$ score.

\item \textbf{External:} ``External'' QE techniques comprise methods that
  obtain expansion terms from other resources besides the target corpus.
  These resources may include other document corpora (including the Web),
  linguistic resources like
  Wordnet\footnote{\url{http://wordnet.princeton.edu}}, and user-query
  logs. \citet{wiki} use Wikipedia\footnote{\url{http://en.wikipedia.org}}
  as a source of expansion terms. Given an initial query, Wikipedia pages
  are retrieved and reranked on the basis of Wikipedia category
  information. The ``best'' wiki pages provide terms for inclusion in the
  expanded query. \citet{xu-sigir09} also used Wikipedia as a source of
  expansion terms. For each query word, the related Wikipedia page (if any)
  is found; terms from this page are ranked, and top-ranked terms are added
  to the query. This approach needs few parameter settings, since for each
  term, only one document is selected.

  \citet{Voorhees94} used Wordnet synsets to find terms related to
  query words. She showed that only the addition of synonyms of query words
  does not consistently improve performance. More recently,
  \citet{fang:2008} showed that Wordnet-based query expansion can yield
  good results if the definitions (or glosses) of words provided by Wordnet
  are used instead of simply relying on the semantic relations defined
  within Wordnet. A comprehensive survey of the uses of ontologies in query
  expansion can be found in \citep{ontology-bhogal}.
\end{itemize}

\subsection{Selective query expansion}
\label{sec:selective}
As mentioned in the Introduction, many of the above QE techniques have been
shown to be effective on the whole over large query sets, even though they
may cause retrieval effectiveness for individual queries to suffer. Our
eventual goal is to formulate a method by which the type of a given query
is first determined, and an appropriate expansion strategy is then used
based on the query category. In other words, we hope to be able to apply QE
techniques tailored to individual queries, rather than applying any
particular QE technique uniformly to all queries. 

As a special case of this problem, researchers have looked at
\emph{selective query expansion}, i.e., the question of whether to expand a
query at all. \cite{selectQE-ecir2004} define an information theoretic
measure that indicates, for a given query, whether it is likely to benefit
from expansion. This measure is used to selectively apply QE to only some
queries. The authors show that their approach works better than applying
QE uniformly across all topics in a test collection.
Similarly, \cite{selQE-cikm2004} show that a comparison between language
models constructed on the basis of the results retrieved by the unexpanded
and a given expanded query can be used to predict whether expansion has
resulted in altering the sense of the original query. In such cases, QE
should be avoided. This idea was shown to be effective in improving the
robustness of expansion strategies.

\section{Query types}
\label{cha:query-types}
As discussed in Section~\ref{sec:query-classification}, queries may be
classified into a wide variety of query types. Thus far, customising online
advertising and search result presentation has been the main motivation
behind query classification: search engines may tailor the format of the
results page or the advertisements displayed in response to a query
according to its category. Our goal in this study is to focus on query
types from a QE perspective. In other words, we are interested in
classification criteria that are likely to have some relation to query
expansion. The types we consider are not mutually exclusive. Our intention
is that the retrieval system (or the user) will decide the (possibly
multiple) categories that a particular query belongs to, and then select
the appropriate QE method for these categories.

Table~\ref{qcategories} lists the query categories that we are interested
in, along with very brief descriptions. Some of these categories can be
determined automatically, while for some, a user's inputs may be required
(these categories are marked M). In some cases, it may be difficult to
categorise queries before an initial retrieval (and evaluation). For
example, to know if a query is hard or not, we need to examine the initial
retrieval results. Generally, we need to expand the queries only if we are
not satisfied by the initial retrieval. The following sections discuss
these categories in more detail. 

\begin{table}[h]
  \begin{center}
    \begin{tabularx}{\textwidth}{ @{} c r X @{}}\toprule
      \textbf{No.} & \textbf{Name} & \textbf{Characteristics} \\\otoprule
      1 & Short query & Few query terms\\ 
      2 & Hard query & Low average precision \\ 
      3 & Ambiguous query & Meaning of query not clear\\ 
      4 & Query containing negative terms & Presence of negation \\ 
      5 & Query involving named entities & Named entities in query \\ 
      6 & Multi-aspect query & Query contains multiple sub-topics \\  
      7 & High-level query & Query uses abstract terms \\ 
      8 & Recall-oriented query (M) & User requires all/many relevant documents \\
      9 & Context implicit in query (M) & Meaning of query determined by
      context \\ 
      10 & Domain specific query (M) & Related to a particular domain \\ 
      11 & Query needing short answer (M) & Specific answer needed \\ 
      13 & Query needing special processing (M) & Special indexing
      techniques may be required \\ 
         & Multilingual query (M) & Query uses more than one language \\ 
         & Noisy query (M) & Query contain some textual error \\ \bottomrule
         
    \end{tabularx}
  \end{center}
  \caption{Query categories} 
  \label{qcategories}
\end{table}

\subsection{Short / long queries}
\label{sec:short-long}
A query may be classified as short or long based on the number of terms or
keywords that it contains. In order to make this notion concrete, we adopt
the following definitions.
\begin{itemize}
\item \emph{short} queries: queries containing fewer than \textbf{four} words
\item \emph{long} queries: queries containing more than \textbf{ten} terms
\end{itemize}
These definitions may be regarded as rather arbitrary; however, they are
only intended to be indicative. If a query consists of a single named
entity that is four words long, it should really be regarded as a short
query.

It is generally believed that casual users tend to formulate short queries,
while more experienced or professional searchers formulate longer queries
that better represent their information need. Queries provided by various
test collections (e.g., TREC ``topics'')\footnote{Appendix A gives an
  overview of the datasets provided by TREC.} usually have both a short and
a long version. They typically consist of a \emph{title}, a
\emph{description} and a \emph{narrative}. The title fields of these
queries are short, since they are mostly intended to model queries created
by casual users; the descriptions are longer. The Narrative section is only
intended to provide a detailed specification of what the user deems
relevant; it should generally not be used as a source of keywords. Table
~\ref{querylength} shows the maximum and minimum lengths (in words) of
different parts of TREC queries.

\begin{table}[h]
\begin{center}
  \begin{tabu} to 0.8\linewidth {c X[1,c] c c}\toprule
    Number & Query field & Maximum length & Minimum length \\\otoprule
    1 & Title & 21 & 1 \\
    2 & Desc & 46 & 5 \\
    3 & Narr\linebreak\small{(400 queries have narr)} & 129 & 14 \\\bottomrule
  \end{tabu}
\end{center}
\caption{TREC queries 1-450: query length in words} 
\label{querylength}
\end{table}

Table~\ref{trecquery} shows the distribution of the length of the
\emph{title} field of TREC queries 1--450 (queries 201--250 are omitted
from this table since they do not contain a title field). We can see from
the table that more than half the queries contain no more than 3 words.
Only occasionally are they any longer, for example, when the title contains
some well known phrase or a long proper name.

\begin{table}[h]
  \begin{center}
    \begin{tabular}{c c}\toprule
      Query length & Number of Queries \\\otoprule
      1 & 18 \\ 
      2 & 101 \\
      3 & 113 \\
      4 & 50 \\ 
      5 & 32 \\ 
      6 & 30 \\ 
      $>$6 & 56 \\ \bottomrule
    \end{tabular}
  \end{center}
  \caption{Distribution of length of titles of TREC queries 1-200 and
    251-450}  
  \label{trecquery}
\end{table}

\param{Benefits of expanding short queries.} We now turn to the
relationship between the length of a query and how it may be affected by
QE. Given their brevity, it is reasonably likely that a short query is an
incomplete representation of the user's information need. Expanding a short
query is likely to yield a more complete representation of the user's
information need. Thus, the benefits of QE are expected to be substantial
in the case of short queries.
%
On the other hand, a long
query is usually a more comprehensive statement of the searcher's
information need. A higher level of retrieval effectiveness can generally
be obtained using long queries, and there is less opportunity for QE
techniques to yield dramatic improvements for such queries.

Table~\ref{shrtquery} illustrates these points. It shows the number of
queries for which a standard QE technique results in better / worse
performance. QE improves effectiveness for 98 out of 150 short, title-only
queries (T). The maximum improvement in MAP over all queries is as much as
0.6016. In contrast, QE yields improvements for fewer medium (TD) or long
(TDN) queries; further, the maximum improvement obtained is also
substantially smaller for long queries (about 0.45).

\begin{table}[h]
  \begin{center}
    \begin{tabular}{c c c c c}\toprule
      Query field(s) & MAP            & MAP        & \# queries improved & \# queries hurt    \\
                     & (no expansion) & (after QE) & (best difference)   & (worst difference) \\ \otoprule
      T & 0.2181 & 0.2630 & 98 (0.6016) & 50 (0.3404) \\
      TD & 0.2560 & 0.2693 & 80 (0.5824) & 70 (0.3827) \\
      TDN & 0.2567 & 0.2749 & 79 (0.4537) & 70 (0.3320) \\\bottomrule
    \end{tabular}
  \end{center}
  \caption{Improvements due to QE for short / long queries (Query set:
    TREC678 (queries 301--450), IR system: TERRIER, term-weighting method:
    IFB2c1.0, QE method: Bo1-based pseudo relevance feedback (40 terms from
    top ranked 10 documents))}
  \label{shrtquery}
\end{table}

\param{Risks related to expanding long / short queries.}
Short queries usually contain only important keywords. Users generally do
not include stop-words (words such as articles, conjunctions, prepositions
that have a primarily grammatical function) in short queries. Thus, short
queries are often not grammatically well-formed sentences or phrases, but
this feature is generally an advantage for many QE techniques: all query
terms can be assumed to be informative, and every query term is likely to
be important during expansion. In contrast, long queries may contain
``weak'' (relatively less useful~/ informative) terms in addition to the
important keywords. Two examples from the TREC topic set illustrate the
important differences between long and short queries.
\begin{itemize}
\item Oil Spill (number-154)\\
  Long: A relevant document will note the location of the spill, amount of
  oil spilled, and the responsible corporation, if known. This will include
  shipborne accidents, offshore drilling and holding tank spills, but
  should not include intentional spills such as Iraq/Kuwait or leakage from
  broken pipes. References to legislation brought about by a spill,
  litigation and clean up  efforts associated with a spill are not relevant
  unless specifics of the spill are included.

\item Black Monday (number-105)\\
  Long: Document will state reasons why U.S. stock markets crashed on 19
  October 1987 (``Black Monday''), or report on attempts to guard against
  another such crash.
\end{itemize}
The short version of these search topics (``Oil Spill'', ``Black Monday'')
contain only keywords, but they do not properly describe the user's
information need. In contrast, the long queries contain a clear and
detailed specification of the user's requirement in natural language.
However, they contain a number of unimportant or general terms (e.g.,
relevant, document, note, include, etc.) that would be inappropriate in a
keyword-only version of these queries. At the time of expansion, therefore,
special care is needed in order to identify the strong terms and to avoid
adding words related to weak terms, since this may result in \emph{query
  drift}.

On the other hand, because a short query contains few words, it has a
greater chance of being ambiguous. Compare, for example, the single term
query ``SVM'' with the longer queries ``SVM pattern recognition'' (in which
SVM refers to Support Vector Machines) and ``SVM admission criteria'' (in
which SVM expands to School of Veterinary Medicine). Expanding such a
single-term query by adding words related to the ``wrong'' sense will also
result in query drift. Further, short queries lie outside the scope of QE
techniques that use some form of language analysis. [CITATION???]

\param{Special processing for verbose queries.} Most Web search queries are
also short, being generally 2 or 3 words
long~\citep{Beitzel:2005:IAQ:1106326.1106329}. However, over the last ten
years or so, long, verbose queries are becoming much more frequent. In
2006, Yahoo claimed that 17\% of its queries contained 5 words or
more~\cite{Gupta:2015:IRV:2766462.2767877}. Users create long queries for a
variety of reasons. A number of techniques for processing verbose queries
have been proposed over the years. Many of these focus on automatic methods
for assigning weights to the original query terms that distinguish between
useful terms and weak
terms~\cite{bendersky2008discovering,lease2009improved,bendersky2011parameterized,jia:verbose}.
For a comprehensive overview of these and other approaches to handling
verbose queries, please see \citep{Gupta2015}.



\subsection{Hard queries}
We characterise a query as \emph{hard} if no automatic 
retrieval method yields good performance (as measured by Average Precision
(AP), or by the number of relevant documents initially retrieved, for
example) for the query.

A number of tracks at TREC have focused on hard queries. The goal of the
Robust \linebreak Track~\citep{DBLP:conf/trec/Voorhees03b} (2003--2005) was
to study queries for which performance is generally poor. In 2003, the
topic set for this task consisted of a total of 100 queries. The minimum
and maximum number of relevant documents for any topic was 4 and
 115 respectively.
%
%
The following year (2004), fifty new topics (651--700)
were created for the Robust Track.
%
%
Later, in its final year, the Million Query Track (2007 -- 2009) ~\citep{DBLP:conf/trec/CarterettePFK09} defined hard queries based on the Average Average Precision (AAP) score for a query, which is the average of AP estimates for a single query over all submitted runs. Difficulty levels were automatically assigned to queries by partitioning the AAP score range into three intervals: $[0, 0.06)$ (\emph{hard} queries), $[0.06, 0.17)$ (\emph{medium} queries), and $[0.17, 1.0]$ (\emph{easy} queries). These intervals were chosen so that queries would be roughly evenly distributed. Of the three, the hard category comprises 38\% of all queries, and includes all queries for which no relevant documents were found.

\param{Types of hard queries.}
Hard queries may be grouped into the sub-categories shown in
Table~\ref{tab:hard-types} based on their properties. 

\tabulinesep=4pt
\begin{table}[t]\centering
  \begin{tabu} to 1.0\linewidth {X[l,4,m] X[c,2.5,m] X[c,0.5,m] X[c,0.5,m] X[l,7,m]}\toprule
    \rowfont[l]{\bfseries}
    Category & Examples (TREC query \#) & $~r~$ & $~R$ & Remarks \\\otoprule
    Queries for which there are very few relevant documents &
    \parbox[c]{1cm}{Q303\\Q320\\Q344} &
    10 6 5 &
    10 6 5 & 
    Expanding such queries to target the few ``needles in the haystack'' is
    unlikely to be beneficial in any real sense.\\\midrule

    Queries with several relevant documents, for which recall is reasonably
    high, but \emph{ranking} is poor &
    \parbox[c]{1cm}{Q374\\Q399\\Q435} &
    203 37 44 &
    204 102 117 &
    Since the relevant documents are retrieved at poor ranks, global
    expansion techniques may work better.
    \\\midrule

    Queries with several relevant documents, but for which \emph{recall} is
    poor &
    \parbox[c]{1cm}{Q307\\Q336\\Q389} &
    25 1 3 &
    215 12 194 &
    Automatic expansion techniques are likely to be inappropriate for such
    queries. Manual, interactive expansion may work well.
    \\\bottomrule
  \end{tabu}
  \caption{Types of hard queries with examples from the TREC query
    collection. $R$ denotes the total number of relevant documents for a
    query, and $r$ denotes the number of relevant documents retrieved for
    that query within the top 1000 ranks. CHECK: WHAT SYSTEM?}
  \label{tab:hard-types}
\end{table}

By definition, the initial retrieval results are poor for hard queries. 
In other words, the top retrieved set contains more irrelevant documents
than relevant ones. PRF-based expansion, which assumes that the top-ranked
documents are relevant, is unlikely to work well for such queries, and may
result in severe performance degradation due to query drift. For example,
if we search for information about the TERRIER IR system using only the
term ``terrier'', most / all top retrieved documents may be related to the
breed of dog. Instead of using PRF, adding the terms ``IR'' and ``system''
to the query \emph{manually} is likely to yield definite improvements.

On the other hand, for an easy query (e.g., TREC queries 365, 403 and 423),
the original query terms are generally good enough for retrieving relevant
documents. Thus, most of the desired documents are retrieved early in the
first round, resulting in high AP (AP values for the above queries are
0.8213, 0.8891, 0.7402 resp.). As the user is likely to be satisfied with
the results of the initial retrieval, query expansion should be done in a
fairly conservative way, if at all, i.e., only a small number of terms
(possibly zero) that are strongly related to the original query terms need
be added to the query

For such queries, since the baseline AP
is high, AQE techniques (which may be modelled as having an element of
stochastic error) are more likely to lead to performance degradation.



Of course, while these categories can be defined easily for a TREC-like
test collection, earlier work discussed in Section~\ref{sec:qpp} suggests
that automatically differentiating between these query types is non-trivial
in a real-life setting. The easiest option may be to have the user look at
the initially retrieved set and decide whether a given query is hard or
easy, and accordingly determine whether expansion is needed or not.



\subsection{Ambiguous queries}
\label{sec:ambiguous}
According to WordNet~\citep{Miller:1995:WLD:219717.219748}, the term
\emph{ambiguous} means ``open to two or more interpretations'' or ``of
uncertain nature or significance'' or ``(often) intended to mislead''.
Extending this definition, we can define an ambiguous query as one whose
meaning is not clear, or one which admits of mutilple valid
interpretations. We categorise ambiguous queries into two groups (analogous
to the grouping in \cite{srd-fntir}), which are discussed in the rest of
this section. 

\subsubsection{Queries containing polysemous words}
\label{sec:polysemous}
We first consider queries that are ambiguous because they contain one or
more \emph{polysemous} words, i.e., words that have multiple meanings. For
such queries, a match with a document on an ambiguous term is only weakly
suggestive of relevance, since the term may have been used in a different
sense from the intended one in the matching document. This problem is more
serious if the polysemous word is an important keyword in the query.

Not surprisingly, the TREC query collection contains a number of polysemous
words. A few examples are:
\begin{itemize}
\item TREC query 350 : health and computer \ul{terminals}. The word 
  \emph{terminal} may be used as an adjective; it may also refer to an
  airport terminal.
\item TREC query 355 : ocean \ul{remote} sensing. \emph{Remote} may also
  be used as a noun (as in a ``television remote'').
\item TREC query 397 : automobile \ul{recalls}. \emph{Recall} may be used
  as a verb, or (less commonly) as the name of a metric.
\end{itemize}


\citet{sanderson2008ambiguous} points out that a large class of ambiguous
words viz., words and phrases that are proper nouns, or are used as such,
occur rarely in traditional, TREC-like query collections. The query
``apple'' is a typical example. The word \emph{apple} may refer to the
fruit, or the computer company, or a number of other
entities\footnote{\url{http://en.wikipedia.org/wiki/Apple_(disambiguation)}}.
The term
`Jaguar'\footnote{\url{http://en.wikipedia.org/wiki/Jaguar_(disambiguation)}}
is another typical example. It could refer to the ``big cat'' that is
formally named \emph{Panthera onca}, but it could also refer to other
objects / entities of more recent origin such as cars, bands,
pens,\footnote{\url{http://www.jaguarpen.com}} or one of several
companies\footnote{\url{http://www.jaguarind.com/aboutus/aboutus.html},
  \url{http://www.jaguarltg.com/}}. Acronyms with multiple expansions
(e.g., ``SVM'' discussed in Section~\ref{sec:short-long}), and acronyms
that are also valid words (e.g., FIRE, acronym for Forum for Information
Retrieval Evaluation) constitute another frequently occurring class of
ambiguous queries. These examples show that polysemy in a language
generally increases over time, as new concepts may be tagged with words
from the existing vocabulary. However, these classes of polysemous queries
have not been seriously studied in past research on polysemous queries.

The performance of a system on an ambiguous query depends on the target
collection. Naturally, ambiguity is a concern only if the collection
actually contains the word used in multiple senses. If the word is used in
only one sense in the target collection, then the query is effectively
unambiguous for that collection. This may happen, for example, in
domain-specific search engines (Section~\ref{subsec:domain}).

\bigskip \bigskip
Approaches to handling polysemous query words can broadly be divided into
three groups.

\paras{Word sense disambiguation (WSD).} A very large number of studies
have focused on the general problem of word sense
disambiguation~\citep{wsd-survey}. A significant body of work has also been
done on WSD for IR. Queries may be \emph{explicitly} disambiguated by
tagging each polysemous query word with a sense code which is utilised
when computing query-document scores.

\citet{wsd-pedersen} showed that a word sense disambiguation algorithm can
improve retrieval effectiveness by 7--14\%. Their WSD algorithm was applied
in conjunction with the standard vector space model for IR. The approach
was evaluated using the Category B TREC-1 corpus (WSJ subcollection).

In a later critique, \citet{wsd-Ng} argues that the question of how
effective WSD is for IR remains an unresolved question, with different
researchers reporting contradictory findings. He showed that many studies
that have demonstrated a positive impact of WSD on IR have made use of
small datasets, or weak baselines.
It is generally agreed, however, that polysemy is a more serious
problem for short queries; it is also generally agreed that in situations
where WSD helps IR, an increase in WSD accuracy has a positive impact on IR
effectiveness~\citep{Sanderson2000,wsd-survey}.


\paras{Implicit WSD.} Retrieval methods may also make use of
\emph{implicit} disambiguation methods. For example, consider the query
``erosion of river banks caused during rainy season''. Even though the term
``bank'' is polysemous, a document $D$ that contains the word in its
intended sense is more likely to also contain the terms \emph{erosion},
\emph{river}, or \emph{rain}, as compared to a document $D'$ that uses the
word in the sense of a financial institution. Most reasonable retrieval
models will favour $D$ over $D'$, thus automatically ``selecting'' the
correct sense of \emph{bank}. In other words, the intended sense of a
polysemous word within a long query may be automatically favoured because
of the additional context provided by the other query terms (this point was
also discussed in Section~\ref{sec:short-long}). 

Additional context may also be provided by the earlier queries issued by
the user within the same session. \citet{cao-kdd} use Conditional Random
Fields to model this context, and show that incorporating session
information often improves query disambiguation.

\paras{Search result diversification.} The third approach to handling
ambiguity, specially in the case of short queries, is search result
diversification (SRD)~\citep{srd-fntir}. In SRD, the goal of a system is to
present a result list that contains documents grouped according to the
various possible interpretations of the given query. This allows the user
to select the results corresponding to the appropriate sense of the query.
The user's feedback may be used to expand the query, keeping in mind its
intended sense.



YIPPY\footnote{http://yippy.com} is an example of a real-life search engine
that attempts a form of SRD. It presents a ranked list of links as usual,
but also provides an automatically generated list of ``clouds'', each of
which corresponds to a possible sense of the query term(s).

\param{Expansion of queries containing polysemous words.}
Before expansion, query terms need to be disambiguated, either explicitly
via WSD, or implicitly. Disambiguation is particularly important when
expanding queries using resources like WordNet or Wikipedia. Since these
resources have broad coverage, expansion without prior disambiguation may
result in the inclusion of many terms related to irrelevant senses of the
query term(s). Indeed, the failure of traditional WordNet-based QE
approaches has been attributed to this problem (citation?? Ch.\ Voorhees).
If disambiguation is not possible, then interaction with the user is
needed.

\medskip
Apart from query WSD, WSD may also be applied to documents, but this
practice is computationally expensive, and thus not widespread in practice
[citation??].

\subsubsection{Underspecified queries}
A user strongly focussed on a particular aspect of a topic may be
temporarily oblivious to other aspects of the topic when searching for
information. Thus, the user may not specify which particular aspect related
to the search keyword(s) she is interested in. Alternatively, she may not
be able to think of a precise formulation for her information need on the
spur of the moment, and may provide only a broad specification of the topic
of interest. In such cases, the user's information need may remain unclear
from the query words, even when these words are not polysemous.
Table~\ref{tab:unclear} shows a few possible interpretations of some TREC
queries that are of this kind.


\begin{table}
  \centering
  \begin{tabu} to 1.0\linewidth {X[1,1] X[1,2.5] X[l,7]}\toprule
    \rowfont[l]{\bfseries}
    TREC query \# & Query title & Possible interpretations \\\otoprule
    Q260 &
    Evidence of human life &
    \emph{during a particular period in history}? \linebreak
    in some geographical locations (e.g., desert islands)? \\
    Q364 &
    Rabies &
    \emph{particular cases and corrective action}? \lb
    which animals are carriers? \lb
    signs, symptoms, prevention, treatment? \lb
    overview / encyclopedic entry? \\
    Q376 & 
    mainstreaming & 
    \emph{of children with physical or mental impairments}? \linebreak
    of physically disabled persons in general? \lb
    of tribal / marginalised communities? \lb
    of juvenile delinquents? \\\bottomrule
  \end{tabu}
  \caption{Examples of underspecified queries. The interpretation in
    italics is the one specified in a longer version of the query
    (specifically, in the description field).}
  \label{tab:unclear}
\end{table}

Given that such queries are open to multiple interpretations by humans,
some level of user interaction or true relevance feedback is likely to be
unavoidable in order to obtain satisfactory results from a search engine.
If the documents retrieved in response to the initial query turn out to be
satisfactory, simple PRF is likely to be beneficial (but, of course, no
further QE may be necessary). Otherwise, the best option for the retrieval
system may be to present a diversified set of results (as discussed in
Section~\ref{sec:polysemous}). The user can then provide some feedback by
marking document sets or individual documents, or may simply select one of
the sets, if appropriate.

Recent research has explored the possibility of obtaining implicit feedback
via eye tracking or other neuro physiological signals~\citep{np2,np5}. In a
research setting, this may involve placing potentially intrusive /
bothersome sensors, but with progress, non-intrusive means of obtaining
feedback are likely to emerge. In such cases, a system may be able to
obtain feedback directly from the natural neuro physiological signals
emitted by the user (e.g., her facial expressions), without requiring any
explicit action on her part.

\subsection{Context implicit in queries}
\label{sec:implicit}
Quite often, when a person in a particular situation converts an
information need to an actual query, e.g., ``national elections'', she may
not be consciously aware that the query may have a very different
interpretation for someone in a different situation. Thus, the intent of
such queries becomes clear only when additional information (e.g.,
nationality, gender, location, time at which query was submitted,
demographic information) about the user is known. We refer to this
additional information as \emph{context}; such queries may be termed
\emph{implicit-context} queries. 


\cite{bai2007using} further differentiate between \emph{context around} and
\emph{context within} a query. In their terminology, a user's domain of
interest, her background knowledge and preferences comprise the context
around a query. In this section, we use the word context in this sense. In
contrast, the context within a query refers to the sense-disambiguating
effect of the query words when taken together (as discussed in
Section~\ref{sec:ambiguous} under \textbf{Implicit WSD}). For example,
this ``internal'' context determines that the word \emph{program} in the
query \emph{Java program} is related to the word \emph{computer}, but this
relationship does not hold if the query is \emph{TV program}.
\citeauthor{bai2007using} show how both kinds of context information may be
integrated into a language modeling approach to IR. They report promising
experimental results on the TREC collections.

Table~\ref{tab:context} lists examples of such queries taken from the TREC
query collection. The persons who create the TREC topics are based in the
USA. Thus, the context implicit in Q269 implies that `foreign' means
countries other than the USA. The same query would be interpreted
differently if it were to occur in the CLEF / FIRE / NTCIR query
collections. Since implicit-context queries admit of mutilple valid
interpretations, they are related to ambiguous queries.


\begin{table}
  \centering
  \begin{tabu} to 1.0\linewidth {X[c,1.5] X[l,3] X[l,7]}\toprule
    \rowfont[l]{\bfseries}
    TREC Query \# & Query title & Implicit context \\\otoprule
    269 & Foreign Trade & Location (foreign == countries other than the US) \\
    \bottomrule
  \end{tabu}
  \caption{Examples of queries containing implicit context.}
  \label{tab:context}
\end{table}

Unlike the creators of TREC topics, the overwhelming majority of Web search
engine users are not trained information-seeking professionals. Thus,
implicit-context queries are encountered far more frequently by Web search
engines. In order to improve retrieval effectiveness for such queries,
researchers have focused on personalised
search~\citep{jeh2003scaling,liu2004personalized}, and the use of contextual
information during search~\citep{Coyle2007}. While some systems explicitly
capture or ask for contextual
information~\citep{bharat2000searchpad,glover2001web}, others guess the
context from a user's
actions~\citep{budzik2000user,finkelstein2001placing}, or from query
logs~\citep{huang2003relevant}.




\subsection{Queries involving common nouns or named entities}
\label{sec:common-nouns}
Generally, user-queries contain a significant proportion of
nouns~\citep{lca}. These nouns may be either named entities (NEs) --- names
of persons, places, organisations, etc. --- or common nouns.

\param{Queries containing named entities.}
Many TREC queries contain NEs, e.g., \emph{King Hussain} (Q450), \emph{babe
  ruth} (Q481), \emph{baltimore} (Q478), \emph{Antarctica} (Q353),
\emph{AT\&T} (Q028), and \emph{Smithsonian Institute} (Q686). These are
usually an important (often the most important) component of the query.
Thus, it may generally be assumed that an article should contain the NE in
order to be relevant. Conversely, the presence of the NE in a document is a
reasonable indicator of its relevance. Queries that are focussed on an NE
are often relatively easy. If the query is expanded nevertheless, the
relative importance of the NE with respect to other query terms should be
maintained in the expanded query.

However, if the NE itself is ambiguous (e.g., \emph{Michael Jordan} could
refer to one of several distinct well-known
persons\footnote{\url{https://en.wikipedia.org/wiki/Michael_Jordan_(disambiguation)}}),
then the issues discussed in Section~\ref{sec:ambiguous} need to be
addressed. An additional issue that may arise is the following. A document
containing the NE will usually also contain a number of pronouns referring
to the NE. Anaphora or coreference resolution --- the process of
identifying pronominal references or alternative names for a named entity
--- may therefore be useful.

\param{Queries containing common nouns.}
It may be much harder to obtain satisfactory results if an important aspect
of the query is specified via a common noun. Table~\ref{tab:common-nouns}
shows a few examples of TREC queries belonging to this category.

\begin{table}
  \centering
  \begin{tabu} to 1.0\linewidth {X[l,1.5] X[l,5] X[l,5]}\toprule
    \rowfont[l]{\bfseries}
    TREC query \# & Query title & Possibly relevant snippets \\\otoprule
    Q109 &
    Find Innovative \textbf{Companies} &
    Sony was the first to introduce a video cassette format $\ldots$
    \\
    Q172 &
    The Effectiveness of \textbf{Medical Products} and Related Programs
    Utilized in the Cessation of Smoking. & 
    Nicorette provides nicotine gum and nicotine lozenges to help you quit
    smoking.
    \\
    Q194 &
    The Amount of Money Earned by \textbf{Writers} &
    J.K.\ Rowling has been paid around three quarters of a billion dollars
    by Warner Brothers $\ldots$
    \\\bottomrule
  \end{tabu}
  \caption{Examples of queries containing common nouns.}
  \label{tab:common-nouns}
\end{table}

The words ``Companies'' and ``Writers'' are common nouns. It is entirely
likely that relevant documents for these queries will contain the names of
specific companies or authors, rather than the corresponding common nouns
in their surface forms. Thus, during expansion, such queries should be
handled differently from queries containing named entities. In some cases,
expanding common nouns in the original query using names of specific
instances may be useful. For example, the term `writers' may be expanded by
adding the names of some popular writers. The expanded query should be
appropriately structured, for example, by including the names as a list of
disjuncts along with the term \emph{writer}.

This presupposes access to appropriate ontologies or gazetteer lists that
provide, for example, a list of author or company names. If such resources
are available, it would be more efficient to use these during indexing,
i.e., documents that contain specific author names could be tagged with the
terms \emph{writer} or \emph{author}.

The system also needs to address the additional issue of selecting which
common nouns are to be expanded, since expanding any common noun present in
the query may not be a good idea.

Interestingly, \cite{Buckley:2009:WCI:1644394.1644417} provides an example
of a query that belongs to this category even though it contains an NE.
TREC topic 398 (\emph{Identify documents that discuss the European
  Conventional Arms Cut as it relates to the dismantling of Europe’s
  arsenal.}) turns out to be problematic because the word `Europe' is too
general; relevant documents are likely to discuss moves made by
specific European countries towards disarmament.



TODO: killer bee example



\subsection{Queries containing negative terms}
Sometimes, users may be able to anticipate the types of irrelevant
documents that may be retrieved in response to a given query. In such
situations, a user may want to provide a detailed statement of her
information need that also explicitly specifies what the user is \emph{not}
looking for. Any keywords that are used to characterise irrelevant
information are referred to as \emph{negative terms}.

Consider the query ``terrorist attacks on the US other than 9/11''. Since
the user has explicitly specified that she is not looking for information
about the {9/11} attack, this term should be counted as a negative term for
this query. Likewise, if a user is looking for local restaurants besides
those that serve Chinese food, she may submit ``restaurants not serving
Chinese food'' as her query. For this query, \emph{Chinese} would count as
a negative term. This example is more complex, however, since
\emph{serving} and \emph{food} should probably not be counted as negative
terms, even though the negation qualifies these terms synactically.
Table~\ref{tab:negative} shows some examples of TREC / INEX  queries that
contain negative terms.

\begin{table}[b]
  \centering
  \begin{tabu} to 1.0\linewidth {X[l,2] X[l,4] X[l,9]}\toprule
    \rowfont[l]{\bfseries}
    Query \# & Query title & Narrative \\\otoprule
    TREC Q124 & Alternatives to Traditional Cancer Therapies &
    $\ldots$ any attempt to experiment with or demonstrate the efficacy of
    any non-chemical, non-surgical, or non-\-radiological approach to
    preventing or curing cancer $\ldots$ \\
    INEX Q419 & film starring +"steven seagal" &
    $\ldots$ films played by Steven Seagal, not produced by
    him.
    \\\bottomrule
  \end{tabu}
  \caption{Examples of queries containing negative terms / aspects.}
  \label{tab:negative}
\end{table}


During expansion, queries that contain negative terms need to be handled
carefully. If the negating qualifiers are ignored (as they usually are), QE
is likely to add terms related to topics that are explicitly designated as
irrelevant, leading to a drop in performance. If the negative terms can be
identified, then they may simply be removed from the original query.
A more aggressive approach would be to include the negative terms in a NOT
clause within a structured query. Naturally, for this method to work,
negative terms have to be identified with high accuracy.

To the best of our knowledge, approaches that try to address what the user
does not want have so far focused only on the initial (verbose) queries.
For example, \cite{sp-negative} propose a method to automatically identify
negative terms in verbose queries and to remove them before initial
retrieval. This method is reported to yield improvements across a number of
collections and various retrieval models. We expect that these improvements
will also lead to post-QE improvements.

\subsection{Multi-aspect queries}
A \emph{multi-aspect} query is one that seeks information about a
particular aspect of a broader topic.\footnote{This definition of
  ``multi-aspect'' may appear confusing. However, historically, the broad
  topic and the particular facet of the topic that the user is interested
  in have been regarded as the multiple \emph{aspects} of the
  query\citep{mitra-improving,Buckley:2009:WCI:1644394.1644417}.}
Multi-aspect queries are best understood via examples. Consider the query
``Terrorist attacks on Amarnath pilgrims.'' One could regard ``Amarnath
pilgrims'' as the primary topic of the query. There are various sub-topics
of this general topic: travel routes taken by the pilgrims, places for
pilgrims to stay along the way, etc. In this query, the user is interested
in one specific sub-topic related to Amarnath pilgrims.

TREC query 203, on the economic impact of recycling tires, is a similar
example. The broad topic of this query is recycling, but the user is only
interested in the recycling of \emph{tires} (rather than other material),
and more specifically in the \emph{economic impact} thereof (rather than,
say, the technology involved). Table~\ref{tab:multi} lists a few more
examples of multi-aspect queries from the TREC query set.

Sometimes, a user may designate multiple sub-topics of a topic as
interesting. For a user who is interested in ``causes and effects of
railway accidents'', documents exclusively discussing \emph{either} the
causes \emph{or} the effects of a railway accident are generally regarded as
relevant. Such queries that are ``disjunctive'' in a sense (but possibly
conjunctive in form) have a broader scope than the examples discussed
above, and are expected to be easier to handle. Multi-aspect queries are
usually hard when the multiple aspects are combined in a conjunctive sense.
\citet{Buckley:2009:WCI:1644394.1644417} contains a detailed analysis of
why automatic IR systems frequently find multi-aspect queries hard.

\begin{table}
  \centering
  \begin{tabu} to 1.0\linewidth {X[l,1.5] X[l,6] X[l,3]}\toprule
    \rowfont[l]{\bfseries}
    TREC Query \# & Query title & Aspects \\\otoprule
    Q100 & Controlling the Transfer of High Technology &
    1. High Technology \lb
    2. Transfer \lb
    3. Controlling \\\midrule
    Q294 & Animal husbandry for exotic animals & 
    1. Animal husbandry \lb
    2. exotic animals \\\midrule
    Q299 & Impact on local economies of military downsizing &
    1. military downsizing \lb
    2. local economies \lb
    3. Impact \\\midrule
    Q321 & Women in Parliaments &
    1. Women \lb
    2. Parliaments
    \\\bottomrule
  \end{tabu}
  \caption{Examples of queries containing multiple aspects.}
  \label{tab:multi}
\end{table}

Quite often, AQE methods add terms that are mostly related to the general
topic of the original query (e.g., \emph{recycling} for TREC Q203 discussed
above). This overemphasises one aspect of the query at the expense of the
others, and usually leads to query drift. Ideally, during expansion,
multi-aspect queries should be expanded in a balanced way, i.e., using
terms related to all (or most) of the multiple aspects. This requires
systems to be able to 
\begin{enumerate*}[~(i)]
\item recognise the various aspects of a query, and 
\item to identify which aspect(s) of the query a candidate expansion term
  is related to.
\end{enumerate*}
\citet{mitra-improving} studied some preliminary methods (both manual and
automatic) that try to prevent query drift by ensuring that the query is
expanded in a balanced way. AbraQ, an approach described by
\cite{crabtree2007exploiting}, attempts balanced query expansion in a Web
search setting by first identifying the different aspects of the query,
identifying which aspects are under-represented in the result set of the
original query, and finally, identifying expansion terms that would
strengthen that particular aspect of the query. 

\citet{selectQE-sig2012} also identify ``problematic'' query terms ---
terms that are probably not present in relevant documents --- on the basis
of the term's idf, or by the predicted probability of that term occurring
in the relevant documents. These query terms are selectively expanded. The
final expanded query is a structured query in Conjuctive Normal Form (CNF),
with each conjunct expected to correspond to a query term (or aspect) and
its synonyms. The authors argue that the use of CNF ensures balanced
expansion, minimises topic drift, and yields stable performance across
different levels of expansion.

\cite{wu2012split} propose
a different approach within a \emph{true} relevance feedback
framework that may also be regarded as being targeted towards balanced
expansion. This approach attempts to diversify the set of documents judged
by a user. Instead of simply letting the user judge the top-ranked results
returned in response to the initial query, the system partitions the
initially retrieved documents into sub-lists, and reranks the documents on
the basis of the query term patterns that occur in them (i.e., whether a
document contains only a single term, multiple terms occurring as a phrase,
or in close proximity, etc.). The documents are then presented iteratively
to the user for judgment.

\subsection{``High-level'' query}
Some queries, such as those shown in Table~\ref{hl-list}, contain terms
that correspond to abstract or ``high-level'' concepts. These terms may not
themselves be present in relevant documents; instead, other more concrete
terms may be used to convey specific instances of the same concept.
If one or more such abstract terms form an important component of an
information need, we refer to the corresponding query as a
\emph{high-level} query.

\begin{table}
  \centering
  \begin{tabu} to 1.0\linewidth {X[c,1.5] X[l,4] X[l,6]}\toprule
    \rowfont[l]{\bfseries}
    TREC Query \# & Query title & Abstract concepts \\\otoprule
    142 & 
    Impact of Government Regulated Grain Farming on International Relations &
    Impact; \lb
    Government Regulated; \lb
    (International) Relations. \\\midrule
    352 & 
    British Chunnel impact & impact. \\\midrule
    353 &
    Antarctica exploration & exploration. \\\midrule
    389 & 
    Illegal technology transfer & 
    Illegal (other than peaceful purposes); \lb
    technology transfer (selling their products, formulas, etc.). \\\bottomrule
  \end{tabu}
  \caption{Examples of queries containing abstract or ``high-level'' terms.}
  \label{hl-list}
\end{table}

`Impact' and `effect' are typical examples of such high-level terms.
Consider the query ``effect of tsunami'', for example. Here `effect' is a
high level term, and refers to anything that happened as a result of a
tsunami. A relevant document may not contain the term `effect’; instead, it
may describe the effect of a tsunami using words such as `death toll',
`property damage', etc.

TREC query 389 (``illegal technology transfer'') is another example. The
description field of the query asks: ``What specific entities have been
accused of illegal technology transfer such as: selling their products,
formulas, etc. directly or indirectly to foreign entities for other than
peaceful purposes?'' `Technology transfer' is thus an abstract concept.
Relevant documents may or may not contain this term. Instead, they may
contain terms like `sell', `license', that describe concrete methods of
technology transfer.

\param{Difference with queries involving common nouns
  (Section~\ref{sec:common-nouns}).}
There is a subtle difference between high-level queries and queries
involving common nouns (discussed in Section~\ref{sec:common-nouns}).
Consider an example from Table~\ref{tab:common-nouns}: \emph{writers}. The
`instantiation' of writers, i.e., the set of persons who are writers, is
not dependent on the query context. In contrast, an abstract term may be
instantiated via different sets of keywords, depending on the subject or
domain of the query. The `effects' or `impact' of a natural disaster, a
foreign tour by a head of state, or of substance abuse are likely to be
described using different words. Thus, finding `bag-of-word' equivalents of
such concepts, being context-sensitive, is more difficult. As a result, SEs
often fail to retrieve an adequate number of relevant documents in response
to high-level queries. For the same reason, correctly automatically
expanding such queries is also challenging. \cite{roussinov2010aspect}
shows that external corpora may be mined to obtain words or word sequences
(conditionally) related to high-level query terms. For example, in TREC query
353, the notion of \emph{exploration} may be indicated by the word
\emph{station}, provided it occurs along with the word \emph{Antarctica},
but not as a part of a phrase such as \emph{train station}.

\subsection{Recall-oriented queries} 
\label{sec:recall}
In certain situations, recall is of paramount importance to the user.
Queries issued by a user in such situations can be termed
\emph{recall-oriented}. The TREC million query
track~\citep{DBLP:conf/trec/AllanCDAPK07} defines recall-oriented queries
as ``looking for deeper, more open-ended information whereas
precision-oriented queries are looking for a small, well contained set of
facts''. Some typical recall-oriented search tasks are:\footnote{\url{http://www.isical.ac.in/~fire/2011/slides/fire.2011.robertson.stephen.pdf}}
\begin{itemize}[topsep=0pt,itemsep=0pt]
\item E-discovery: searching for documents required for disclosure in a
  legal case~\citep{oard2010evaluation,oard2013information}. 
\item Prior-art patent search: looking for existing patents which might
  invalidate a new patent application.
\item Evidence-based medicine: finding all prior evidence on treatments for
  a medical case.
\end{itemize}
For these tasks, having to examine several irrelevant documents may be an
acceptable overhead, but the penalty for missing a relevant document is
likely to be high. 

The TREC legal track models a recall-oriented task. Query 100 from this
track reads: ``Submit all documents representing or referencing a formal
statement by a CEO of a tobacco company describing a company merger or
acquisition policy or practice''. Note that the query explicitly requires
\emph{all} relevant documents to be retrieved. This is in contrast to
casual, ad hoc searches, in which users are generally satisfied by a small
number of relevant documents retrieved at the top ranks.

Table~\ref{rcl-exmp} shows that recall generally increases with the number
of terms added to a query during QE. Thus, for recall-oriented queries,
\emph{massive query expansion}, i.e., expansion by adding a very large
number of potentially useful terms that occur in at least one relevant
document, may be a good idea. However, the risk of query drift
significantly increases if massive expansion is based on PRF. Relevance
feedback involving some user interaction may be necessary to ensure high
recall without a concomitant loss in precision. \cite{kripa} have recently
proposed a method that uses the Cluster Hypothesis to effectively leverage
only a modest amount of user interaction for high recall.

\begin{table}[h]
  \centering
  \begin{tabular}{cccc}\toprule
    \#Term & \#rel-ret(among top 1000) &recall@1000 &MAP \\\otoprule
    10 & 8273 & 0.6686 & 0.2452 \\
    20 & 8442 & 0.6784 & 0.2525 \\
    30 & 8530 & 0.6851 & 0.2561 \\
    40 & 8556 & 0.6891 & 0.2574 \\
    50 & 8551 & 0.6901 & 0.2586 \\
    60 & 8562 & 0.6906 & 0.2586 \\
    70 & 8587 & 0.6922 & 0.2595 \\
    80 & 8589 & 0.6927 & 0.2601 \\
    90 & 8602 & 0.6938 & 0.2605 \\
    100 & 8605 & 0.6943 & 0.2611 \\ \hline
  \end{tabular}
  \caption{Effect of increasing the degree of expansion on recall on the
    TREC678 collection
    (expansion method used: KLD, no.\ of top documents: 40).}
  \label{rcl-exmp}
\end{table}

\subsection{Domain specific queries}
\label{subsec:domain}
Queries which are related to and need information from one specific domain
(e.g., sports, medicine, law) are called domain specific queries.
Earlier work on classifying queries according to their domain has been
discussed in Section~\ref{sec:query-classification}. Over the years, TREC
has offered a number of tasks that address IR from specific domains /
genres. Table~\ref{tab:domain} provides a non-exhaustive list of some of
these tasks.

\begin{table}
  \centering
  \begin{tabu} to 1.0\linewidth {X[l,2] X[c,1] X[l,4]}\toprule
    \rowfont[l]{\bfseries}
    Track name & Years & Domain \\\otoprule
    Legal & & \\\midrule
    Enterprise search & & Searching an organisation's data \\\midrule
    Genomics & & Genomics data (broadly construed to include not just gene
    sequences but also supporting documentation such as research papers,
    lab reports, etc. \\\midrule
    Chemical & & Information retrieval and extraction tools for chemical
    literature \\\midrule
    Medical records & & Free-text fields of electronic medical records \\\bottomrule
  \end{tabu}
  \caption{TREC tracks that focus on domain-specific IR.}
  \label{tab:domain}
\end{table}

Domain-specific queries constitute a special case of \emph{Vertical
  search}, where the system caters to users interested in a particular type
of online
content\footnote{\url{https://en.wikipedia.org/wiki/Vertical_search}}. 
Vertical searches may focus not only on a particular domain or topic, but
also on a specific media type or genre of content, e.g., image or video
search, shopping, travel, and scholarly literature.

  

\param{Expansion strategy.} For some domains, it should be possible to
leverage domain-specific lexical resources for expansion. For example, MeSH
or the UMLS metathesaurus may be used to expand queries in the biomedical
domain. \cite{hersh} have reported on the effectiveness of using the UMLS
metathesaurus for QE. Similarly, \citet{mesh} have studied expansion of
PubMed queries using MeSH. Naturally, in order to utilise such
domain-specific ontologies, a system should be able to identify the target
domain of user queries with reasonable accuracy. On the other hand, if the
user explicitly indicates the domain of the query, this not only eliminates
the query-classification step, but should also help to reduce any ambiguity
that might be present. In recent work, \cite{macias2015effects} have
confirmed that the domain of interest is important when quantifying the
semantically relatedness between words. Even though their experiments were
not directly related to QE, their findings are expected to be applicable
when estimating the semantic relation between the query and candidate
expansion terms.





\subsection{Short answer type queries}
Some queries need very specific and `to the point' answers that comprise a
few words, a single sentence, or a short passage. In such cases, the user
does not want to read a full document, or a long passage to find the answer.
%
Most queries starting with `what', `who', `where', `when', `which', `whom',
`whose', `why' etc.\ fall in this category. There are few examples of
such queries in the TREC adhoc dataset, but the query sets for the
Question-Answering (QA) tasks at TREC, CLEF and NTCIR consist of these types
of queries.

Systems that effectively address such queries usually have the following
architecture~\citep{Prager:survey}. The question is first analysed to determine 
the answer type, and to generate an appropriate keyword query. The keyword
query is used to retrieve a set of passages (or documents) from a corpus. The
retrieved passages are analysed to generate a list of candidate answers.
The candidate answers are further processed to generate the final ranked
list of answers.

Query expansion can, and often does, play a role in retrieving passages or
documents in response to the keyword query. In one of the best-known QA
systems~\citep{pasca2001high,Moldovan2003}, some of the question words are
selected as keywords (using mainly part of speech information). The
original question is parsed to determine dependencies between the question
words, which are in turn used to order the list of selected keywords. These
keywords are also spell-checked; spelling variants are added to the query
if necessary. The most important of these keywords are used to retrieve
documents using the Boolean model. From these documents, the system
extracts paragraphs or smaller text passages containing all keywords in
close proximity of one another. If too many paragraphs are retrieved, the
query is expanded by including additional terms from the list of keywords;
if too few paragraphs are retrieved, some of the keywords from the initial
query are dropped. The system also employs QE in a more traditional way by
using WordNet to expand the query keywords with morphological, lexical and
semantic alternatives.

\subsection{Queries that need special handling during query processing}
\label{sec:preproc}
Query processing generally includes some (or all) of the following steps:
stopword removal, stemming, case normalisation, treatment of acronyms and
numbers, handling spelling errors, etc. In this section, we consider
queries that need special handling during query processing, i.e., queries
for which the general (or ``standard'') query processing methods would
result in a loss of some important information, which in turn would lead to
poor retrieval effectiveness. Note that this special processing must be
done on the initial query; the question of whether (or how) to expand the
query arises later. Indeed, without this special processing, initial
retrieval effectiveness may be so poor that any subsequent expansion of the
query would be pointless.

\begin{itemize}
\item \textbf{Stopword removal.} Articles, conjunctions,
  prepositions and other frequently occurring words are discarded as
  stopwords because they usually have a grammatical function, and are not
  indicative of the subject matter of documents and queries. The words
  \emph{before} and \emph{after} are two examples of such words that are
  included in the default stopword list used by TERRIER. However, in a
  query like ``increased security measures after 9/11'', the word ``after''
  is an important qualifier. Discarding it as a stopword during indexing of
  documents and queries may cause problems.

\item \textbf{Case normalisation.} Many IR systems reduce all alphabets to
  their lowercase forms during indexing.
  Since proper nouns can be identified by a starting capital letter, this
  case normalisation may result in loss of information in some cases. In
  TREC query 409 (\emph{legal, Pan Am, 103}), the word \emph{Am} is not
  actually used as a stopword; however, through a combination of case
  normalisation and stopword removal, many systems would incorrectly
  discard this word from the query. This problem would also arise if the
  acronym \emph{U.S.}\ were written as \emph{US}.

  The Smart system~\cite{} ran into a related problem during the initial
  years of TREC: because the ampersand in query 028 (\emph{AT\&T's
    Technical Efforts}) was treated as a word delimiter, \emph{AT\&T's} was
  tokenised as \emph{at}, \emph{\&}, \emph{t}, and \emph{'s} after case
  normalisation, and all four tokens were discarded, resulting in very poor
  performance for this query. This problem might also occur with TREC query
  391 (\emph{R\&D drug prices}), with \emph{R\&D} being tokenised as
  \emph{r} and \emph{d}.


\item \textbf{Identification of numbers.} A user query may contain numbers
  denoting a year, a flight number or something similar which is an
  integral part of the information need. Simply ignoring numbers during
  indexing of either documents or queries (such as TREC query 409 discussed
  above) may have a significant detrimental effect.

\item \textbf{Stemming.} Stemming is used to conflate morphological
  variants of a word to a canonical form, so that a keyword in a query
  matches a variant occurring in a document. Whether a query word should be
  stemmed or not often depends on the query, and more specifically the
  sense of the query word. For example, in a query about Steve Jobs, the
  word `Jobs' should not be stemmed to `job'. Similarly, the word `apples'
  occurring in a document about the fruit should not be stemmed to match
  the word `Apple' in a query about Apple's marketing strategy for the
  iPhone. \cite{paik2013effective} show that a query-specific stemming
  approach is significantly more effective than applying a generic stemmer
  uniformly to all queries and documents in a collection. To achieve this
  effect, documents should not be stemmed at the time of indexing. Instead,
  a given query should be expanded by adding to it only the \emph{desirable}
  variants of query keywords. 


\item \textbf{Indexing phrases.} The question of whether to use phrases ---
  multiple words that occur contiguously or in close proximity and
  constitute a single semantic unit, e.g., blood cancer, machine learning
  --- during indexing and retrieval has been investigated in a number of
  studies~\citep{fagan1987experiments,mitra1997analysis}. This question is
  also tied to the issue of whether to use phrases during QE. The use
  of phrases has been found to generally improve performance, though its
  effect is not always significant. \cite{song2006keyphrase} show that
  keyphrases extracted from retrieved documents may be useful as expansion
  terms. Their keyphrase extraction algorithm makes use of the occurrences
  of stopwords in the documents. Thus, in order to use their method in a
  practical SE, documents and queries need special handling during indexing
  and retrieval.

\end{itemize}

\subsubsection*{Multi-lingual query}
Multilingual queries, i.e., queries that make use of words from more than
one language (say, $L_1, L_2, \ldots, L_k$), are a particular class of
queries that need special handling. Such queries~\citep{multq} are common
in multilingual countries or communities like India or the EU. A number of
factors lead to the creation of multilingual queries.
\begin{itemize}
\item The amount and variety of native language content on the Web is still
  rather low for many languages, e.g., Assamese or Punjabi. An Assamese
  user may be able to read English fluently, and is thus likely to know the
  most important English keywords related to her information need. At the
  same time, she may be unable to find appropriate English words to
  completely formulate her query in English. For example, consider a user
  who is looking for the differences between interpreters and assemblers.
  For such a user, it would be natural to submit a query that mixes the
  English words \emph{interpreters} and \emph{assemblers}) with the
  Assamese equivalents of \emph{difference} and the remaining words.
\item In a country like India, where the language used at work is often
  English, users may not be familiar with the local equivalents of all
  technical terms. If such a user is specifically interested in a technical
  or official document in her native language, her natural tendency would
  be to search using a mix of English and native language words.
\item Some English terms are very commonly used in non-English-speaking
  regions. For example, in Bengali documents, the term `recipe' is more
  likely to be used than the Bengali equivalent 
  (\emph{randhanpronali}). In addition, documents may use either the
  Bengali transliteration of `recipe', or the original Roman form of the
  word. An experienced user who is aware of this may include all three
  terms in her query for better recall.
\end{itemize}

  
When processing a multilingual query, a system needs to address the
following problems.
\begin{itemize}
\item \textbf{Source language identification.} If words from multiple
  languages are present in the query, then their respective languages have
  to be identified. This is trivial if the languages use distinctive
  scripts, but if any of the languages involved shares its script with
  other languages, the language identification problem becomes harder. If
  different inverted indices are maintained for different languages, the
  system also needs to determine which target collections need to be
  searched for a given multilingual query.
\item \textbf{Transliteration.} For a very long time, native language
  keyboards were a rarity for many languages. Users of these languages were
  habituated to using the Roman script when writing in their language. Such
  habits die hard, and many users continue to prefer using the Roman script
  to write in their language. In order to retrieve documents in the
  original language, the system needs to first back-transliterate words
  from Roman to the native language.

  Moreover, if a query entered by such a user is multilingual, word-level
  language identification may be harder, since the Roman script is used for
  all words. The problem is compounded further if, after transliteration,
  words in the user's native language match valid English words. For
  example, \emph{More} is both a valid English word, and a reasonably
  common surname in Marathi. Similarly, \emph{Shulk} is a fictional
  character and the main protagonist in a popular video game; it also means
  \emph{tax} in Hindi and other Indian languages.
\end{itemize}
Some of these problems are being studied within the ``Search in the
Transliterated Domain'' track at FIRE. This track involves two subtasks:
\begin{enumerate*}[~(i)]
\item given a multilingual query, label each word with its language; and
\item reverse-transliterate non-English words written in Roman script into
  their native script.
\end{enumerate*}
If these problems are not properly addressed, QE may hurt performance. If
an important expansion term in one language happens to be a valid word in
another language, the system also needs to carefully consider the net
benefit of including such an ambiguous term in the expanded query when
retrieving documents from a multilingual collection.



   

\subsubsection*{Noisy queries}
\label{sec:noisy}
Finally, we consider queries that need special handling because of the
presence of errors or noise. Such noise can be introduced because of
spelling errors committed by the user, or because queries are submitted via
a noise-inducing interface, e.g. spoken queries, querying via mobile
messaging, and queries written by hand using a stylus.

Since queries in most test collections are methodically created by
experienced or professional users of IR systems, such queries are usually
free from noise. TREC query 464 --- \emph{nativityscenes} --- is one of the
rare TREC queries that contain an error. However, query logs of practical
search engines are likely to have large numbers of such examples.

Noisy queries also need special handling, usually spelling correction. A
fair amount of work has recently been done on spelling correction in
queries~\cite{gao2010large,duan2011online,li2012generalized}, and a large
number of patents exist for such techniques. Some of these methods are
employed in many practical Web search engines that are often able to
suggest or even automatically provide corrections for such noisy queries.






\section{Conclusions and future work}
\label{cha:conclusions}
Query expansion is a standard technique for addressing the well-known
vocabulary mismatch problem faced by IR systems. Over the years, a number
of effective QE techniques have been proposed. However, the effect of
different QE techniques for individual queries can vary greatly.

Our long-term goal is to improve overall performance by applying QE
techniques tailored to a given query, rather than applying the same general
QE method to all queries. To this end, we have proposed a taxonomy of query
classes. Not all proposed query categories are new. However, we have
specifically considered query categorisation from a QE perspective.

We have discussed the properties of each query class with examples. We have
also proposed some QE strategies that might be effective for each query
category. We believe that there is significant scope for future work in a
careful investigation of the most effective QE techniques for each query
class.

Our next step will be to come up with more precise formulations of QE
techniques for the various categories and to test these proposed techniques
using standard datasets. While for many query categories, such testing can
be done using TREC datasets, a few categories pertain specifically to Web
queries. 

An additional challenge will be to automatically detect the type of a given
query. This is likely to be straightforward for some query types, but we
will need to systematically study automatic query classification approaches
in future work.

To conclude, we believe that in the recent future, as the Web continues to
grow, and search becomes a more and more frequent activity, IR systems will
need customised methods for individual queries and users. The work
described in this report is an initial step in this direction. 


\appendix
\section{TREC: Text REtrieval Conference}



Exerimental IR, like any other experimental discipline, depends heavily on
the existence of standardised benchmark datasets, or \emph{test
 collections}. A test collection in IR is a collection of documents
along with a set of test queries. The set of relevant articles for 
each query is also known. To measure the effectiveness of a technique,
documents are retrieved using that technique for each test query in the
collection. Using the relevance information for the queries, the average
precision value can be computed for each query. The mean average precision
for the entire query set is then calculated. Different techniques can be
compared using the average precision figures they yield on a given test
collection. Obviously, techniques that perform well across a wide variety
of test collections can be regarded as robust. 

For our experiments, we use parts of the TREC
collection~\citep{trec-book}. TREC (Text REtrieval Conference) is an ARPA
and NIST co-sponsored effort that brings together information retrieval
researchers from around the world to discuss their systems and to evaluate
them on a common test platform. The documents and queries in this
collection are described below. 

\subsection{Documents}
\label{sec:documents}
The TREC document collection consists of a large number of
full-text documents drawn from a variety of sources. The documents are
stored on CD-ROMS, called the TREC disks. The disks are numbered, and a
combination of several disks can be used to form a text collection for
experimentation. Some statistics about the data on various disks is listed
in Table~\ref{docstats} (adapted from \citep{Harman-TREC6}). The sources
for the data are: 
\begin{itemize}
\item Disk 1
 \begin{itemize}
 \item AP Newswire, 1989. (AP)
 \item Short abstracts from the U.S.\@ Department of Energy publications. (DOE)
 \item U.S.\@ Federal Register, 1989. (FR)
 \item Wall Street Journal, 1987--1989. (WSJ)
 \item Articles from {\em Computer Select\/} disks, Ziff Davis 
 Publishing. (ZIFF) 
 \end{itemize}
\item Disk 2
 \begin{itemize}
 \item AP Newswire, 1988. (AP)
 \item U.S.\@ Federal Register, 1988. (FR)
 \item Wall Street Journal, 1990--1992. (WSJ)
 \item Articles from {\em Computer Select\/} disks, Ziff Davis
 Publishing. (ZIFF). 
 \end{itemize}
\item Disk 3
 \begin{itemize}
 \item AP Newswire, 1990. (AP)
 \item U.S.\@ Patents, 1993. (PAT)
 \item San Jose Mercury News, 1991. (SJMN)
 \item Articles from {\em Computer Select\/} disks, Ziff Davis
 Publishing. (ZIFF)
 \end{itemize}
\item Disk 4
 \begin{itemize}
 \item Financial Times, 1991--1994. (FT)
 \item U.S.\@ Federal Register, 1994. (FR)
 \item U.S.\@ Congressional Record, 1993. (CR)
 \end{itemize}
\item Disk 5
 \begin{itemize}
 \item Foreign Broadcast Information Service. (FBIS)
 \item LA Times. (LAT)
 \end{itemize}
\end{itemize}

\begin{table}
\begin{center}
\caption{TREC Document Statistics \label{docstats}}
\ \\
{\small
\begin{tabular}{|c||c|c|c|c|} \hline
Source & Size (Mb) & Number of articles & Median number & Average number \\
& & & of terms/article & of terms/article \\ \hline \hline
\multicolumn{5}{|c|}{Disk 1} \\ \hline
WSJ & 270 & 98,732 & 182 & 329 \\ \hline
AP & 259 & 84,678 & 353 & 375 \\ \hline
ZIFF & 245 & 75,180 & 181 & 412 \\ \hline
FR & 262 & 25,960 & 313 & 1017 \\ \hline
DOE & 186 & 226,087 & 82 & 89 \\ \hline \hline
\multicolumn{5}{|c|}{Disk 2} \\ \hline
WSJ & 247 & 74,520 & 218 & 377 \\ \hline
AP & 241 & 79,919 & 346 & 370 \\ \hline
ZIFF & 178 & 56,920 & 167 & 394 \\ \hline
FR & 211 & 19,860 & 315 & 1073 \\ \hline \hline
\multicolumn{5}{|c|}{Disk 3} \\ \hline
SJMN & 290 & 90,257 & 279 & 337 \\ \hline
AP & 242 & 78,321 & 358 & 379 \\ \hline
ZIFF & 349 & 161,021 & 119 & 263 \\ \hline
PAT & 245 & 6,711 & 2896 & 3543 \\ \hline\hline
\multicolumn{5}{|c|}{Disk 4} \\ \hline
FT & 564 & 210,158 & 316 & 413 \\ \hline
FR94 & 395 & 55,630 & 588 & 645\\ \hline
CR & 235 & 27,922 & 288 & 1374 \\ \hline\hline
\multicolumn{5}{|c|}{Disk 5} \\ \hline
FBIS & 470 & 130,471 & 322 & 544 \\ \hline
LAT & 475 & 131,896 & 351 & 527 \\ \hline
\end{tabular}
}
\end{center}
\end{table}

\subsection{Queries}
\label{sec:queries}
The queries are natural-language queries supplied by users. Most queries
consist of 3 parts:
\begin{itemize}
\item \emph{Title}: a few keywords (usually 2--3) related to the user’s query,
\item \emph{Desc} (description): a short, natural-language statement of the
  user's information need,
\item \emph{Narr} (narrative): a more detailed specification of what makes
  a document relevant for the corresponding topic.
\end{itemize}

The queries have varied widely from year to year. At the first two
conferences, TREC--1 and TREC--2, the queries were quite long and
represented long-standing user information needs. Reflecting a trend
towards realistic user queries, the queries for TREC--3 were considerably
shorter and the queries for TREC--4 were just a sentence or two. Some
characteristics of the query sets are shown in Table~\ref{query-stats} (the
training queries were provided to help train systems for TREC--1). 

Users also provide relevance judgments (i.e. they specify which documents
are useful and which are non-relevant) for the documents in the
collection. These judgements enable us to measure the retrieval
effectiveness (using average precision figures) of our algorithms. 

\begin{table}
\begin{center}
\caption{Query Statistics \label{query-stats}}
\ \\
{\small
\begin{tabular}{|c|c|c|c|c|} \hline
 Query Id. & \# of Queries & Min & Max & Mean \\\hline
 TREC--1 & 50 & 44 & 250 & 107.4 \\
 51--100 & & & &\\\hline
 TREC--2 & 50 & 54 & 231 & 130.8 \\
 101--150 & & & &\\\hline 
 TREC--3 & 50 & 49 & 180 & 103.4 \\
 151--200 & & & &\\\hline 
 TREC--4 & 50 & 8 & 33 & 16.3 \\
 201--250 & & & &\\\hline 
 TREC--5 & 50 & 29 & 213 & 82.7 \\
 251--300 & & & &\\\hline
 TREC--6 & 50 & 47 & 156 & 88.4 \\
 301--350 & & & &\\\hline
 TREC--7 & 50 & 31 & 114 & 57.6 \\
 350--400 & & & &\\\hline
 TREC--8 & 50 & 23 & 98 & 51.8 \\
 401--450 & & & &\\\hline

\end{tabular}
}
\end{center}
\end{table}

\bigskip

In recent years, the TREC collection has emerged as a standard test
collection for experimental IR. At TREC--6, the sixth in this series of
conferences, thirty-eight groups including participants from nine different
countries and ten companies were represented. Given the participation by
such a wide variety of IR researchers, a large and heterogeneous collection
of full-text documents, a sizeable number of user queries, and a set of
relevance judgments, TREC has rightfully become a standard test
environment for current information retrieval research. 

\bibliographystyle{plainnat}
\bibliography{qc}

\end{document}